\documentclass[prl,aps,amsfonts,amssymb,amsmath,floats,floatfix,twocolumn,twocolumn,superscriptaddress]{revtex4-1}

\usepackage{graphicx}
\usepackage{dcolumn}
\usepackage{bm}

\begin{document}

\title{Evidence for superconductivity in Li-decorated monolayer graphene} 

\author{B.M. Ludbrook}
\affiliation{Department of Physics {\rm {\&}} Astronomy, University of British Columbia, Vancouver, British Columbia V6T\,1Z1, Canada}
\affiliation{Quantum Matter Institute, University of British Columbia, Vancouver, British Columbia V6T\,1Z4, Canada}
\author{G. Levy}
\affiliation{Department of Physics {\rm {\&}} Astronomy, University of British Columbia, Vancouver, British Columbia V6T\,1Z1, Canada}
\affiliation{Quantum Matter Institute, University of British Columbia, Vancouver, British Columbia V6T\,1Z4, Canada}
\author{P. Nigge}
\affiliation{Department of Physics {\rm {\&}} Astronomy, University of British Columbia, Vancouver, British Columbia V6T\,1Z1, Canada}
\affiliation{Quantum Matter Institute, University of British Columbia, Vancouver, British Columbia V6T\,1Z4, Canada}
\author{M. Zonno}
\affiliation{Department of Physics {\rm {\&}} Astronomy, University of British Columbia, Vancouver, British Columbia V6T\,1Z1, Canada}
\affiliation{Quantum Matter Institute, University of British Columbia, Vancouver, British Columbia V6T\,1Z4, Canada}
\author{M. Schneider}
\affiliation{Department of Physics {\rm {\&}} Astronomy, University of British Columbia, Vancouver, British Columbia V6T\,1Z1, Canada}
\affiliation{Quantum Matter Institute, University of British Columbia, Vancouver, British Columbia V6T\,1Z4, Canada}
\author{\\D.J. Dvorak}
\affiliation{Department of Physics {\rm {\&}} Astronomy, University of British Columbia, Vancouver, British Columbia V6T\,1Z1, Canada}
\affiliation{Quantum Matter Institute, University of British Columbia, Vancouver, British Columbia V6T\,1Z4, Canada}
\author{C.N. Veenstra}
\affiliation{Department of Physics {\rm {\&}} Astronomy, University of British Columbia, Vancouver, British Columbia V6T\,1Z1, Canada}
\affiliation{Quantum Matter Institute, University of British Columbia, Vancouver, British Columbia V6T\,1Z4, Canada}
\author{S. Zhdanovich}
\affiliation{Quantum Matter Institute, University of British Columbia, Vancouver, British Columbia V6T\,1Z4, Canada}
\affiliation{Department of Chemistry, University of British Columbia, Vancouver, British Columbia V6T\,1Z1, Canada}
\author{D. Wong}
\affiliation{Department of Physics {\rm {\&}} Astronomy, University of British Columbia, Vancouver, British Columbia V6T\,1Z1, Canada}
\affiliation{Quantum Matter Institute, University of British Columbia, Vancouver, British Columbia V6T\,1Z4, Canada}
\author{P. Dosanjh}
\affiliation{Department of Physics {\rm {\&}} Astronomy, University of British Columbia, Vancouver, British Columbia V6T\,1Z1, Canada}
\affiliation{Quantum Matter Institute, University of British Columbia, Vancouver, British Columbia V6T\,1Z4, Canada}
\author{\\C. Stra{\ss}er}
\affiliation{Max Planck Institute for Solid State Research, 70569 Stuttgart, Germany}
\author{A. St{\"o}hr}
\affiliation{Max Planck Institute for Solid State Research, 70569 Stuttgart, Germany}
\author{S. Forti}
\affiliation{Max Planck Institute for Solid State Research, 70569 Stuttgart, Germany}
\author{C.R. Ast}
\affiliation{Max Planck Institute for Solid State Research, 70569 Stuttgart, Germany}
\author{U. Starke}
\affiliation{Max Planck Institute for Solid State Research, 70569 Stuttgart, Germany}
\author{A. Damascelli}
\email{damascelli@physics.ubc.ca}
\affiliation{Department of Physics {\rm {\&}} Astronomy, University of British Columbia, Vancouver, British Columbia V6T\,1Z1, Canada}
\affiliation{Quantum Matter Institute, University of British Columbia, Vancouver, British Columbia V6T\,1Z4, Canada}


\maketitle 

{\bf
Monolayer graphene exhibits many spectacular electronic properties, with superconductivity being arguably the most notable exception. It was theoretically proposed that superconductivity might be induced by enhancing the electron-phonon coupling through the decoration of graphene with an alkali adatom superlattice [Profeta \textit{et al.} \textit{Nat. Phys.} \textbf{8}, 131-134 (2012)]. While experiments have indeed demonstrated an adatom-induced enhancement of the electron-phonon coupling, superconductivity has never been observed. Using angle-resolved photoemission spectroscopy (ARPES) we show that lithium deposited on graphene at low temperature strongly modifies the phonon density of states, leading to an enhancement of the electron-phonon coupling of up to $\lambda\!\simeq\!0.58$. On part of the graphene-derived $\pi^*$-band Fermi surface, we then observe the opening of a $\Delta\!\simeq\!0.9$\,meV temperature-dependent pairing gap. This result suggests for the first time, to our knowledge, that Li-decorated monolayer graphene is indeed superconducting with $T_c\!\simeq\!5.9 K$.
}

While not observed in pure bulk graphite, superconductivity occurs in certain graphite intercalated compounds (GICs), with $T_c$ of up to 11.5\,K in the case of CaC$_6$ \cite{Emery2005,Weller2005}. The origin of superconductivity in these materials has been identified in the enhancement of electron-phonon coupling induced by the intercalant layers \cite{Mazin2005a,Calandra2005a}. The observation of a superconducting gap on the graphitic $\pi^*$  bands in bulk CaC$_6$ \cite{Yang2014} suggests that realizing superconductivity in monolayer graphene might be a real possibility. This has indeed attracted intense theoretical and experimental efforts \cite{Uchoa2007,McChesney2010,Profeta2012,Nandkishore2012,Guzman2013,Fedorov2014,Margine2014}. In particular, recent density-functional theory calculations have suggested that, analogous to the case of intercalated bulk graphite, superconductivity can be induced in monolayer graphene through the adsorption of certain alkali metals \cite{Profeta2012}. 

Although the Li-based GIC -- bulk LiC$_6$ -- is not known to be superconducting, Li decorated graphene emerges as a particularly interesting case with a predicted superconducting $T_c$ of up to $8.1$\,K \cite{Profeta2012}. The proposed mechanism for this enhancement of $T_c$ is the removal of the confining potential of the graphite C$_6$ layers, which changes both the occupancy of the Li $2s$ band (or the ionization of the Li) and its position with respect to the graphene layer. This ultimately leads to an increase of the electron-phonon coupling constant from $\lambda\!=\!0.33$ to 0.61, in going from bulk to monolayer LiC$_6$. It has been argued that the LiC$_6$ monolayer should exhibit the largest values of both $\lambda$ and $T_c$ among all alkali-metal-C$_6$ superlattices \cite{Profeta2012}. Nevertheless, while there is thorough experimental evidence for adatom-enhanced electron-phonon coupling in graphene \cite{McChesney2010,Bianchi2010,Fedorov2014}, superconductivity has not yet been observed in decorated monolayer graphene.

\begin{figure*}[t!]
\includegraphics[width=0.85\textwidth]{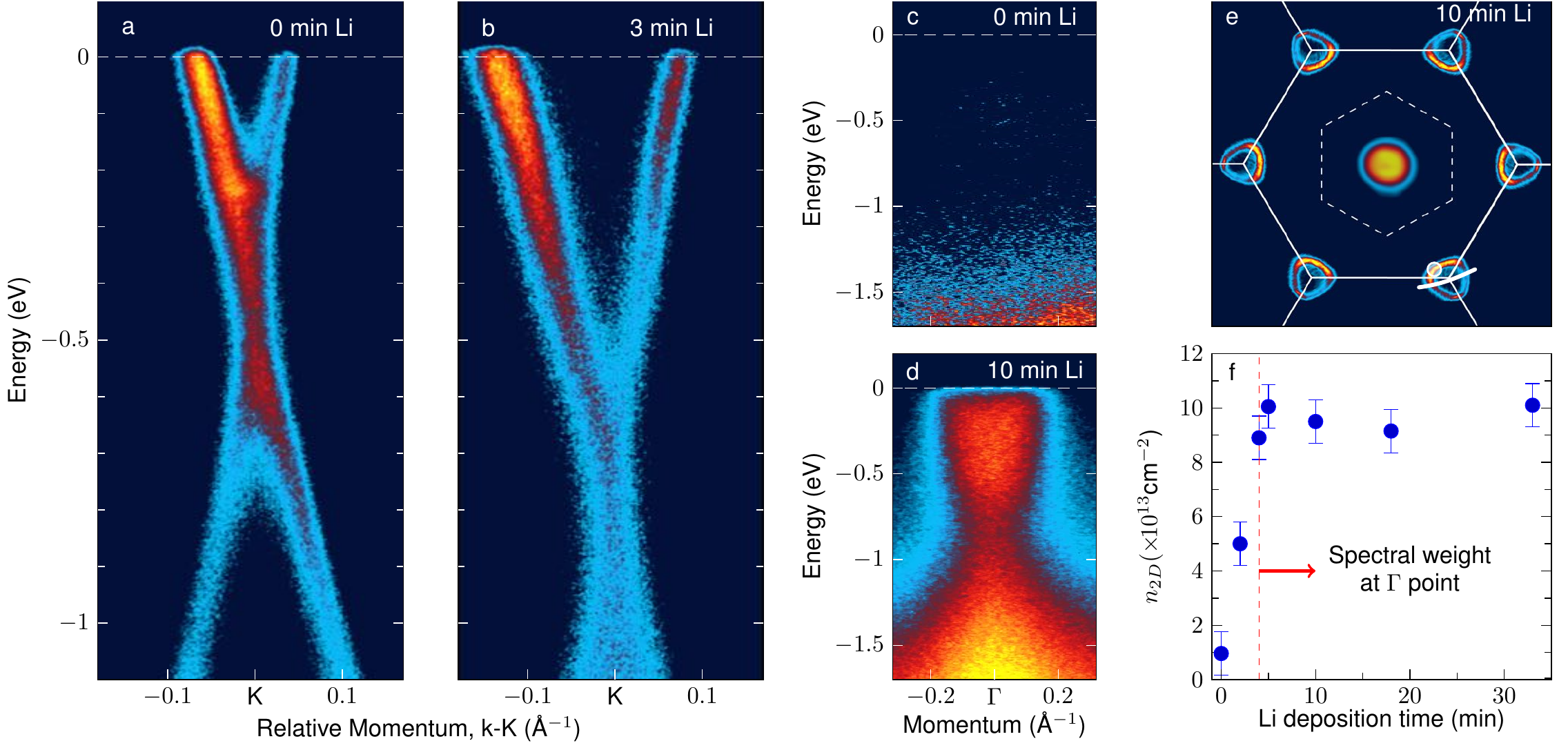}
\caption{\label{Fig:1}\textbf{Charge-transfer doping of graphene by lithium adatoms.} (a) Dirac-cone dispersion measured by ARPES at 8\,K on pristine graphene and (b) after 3 minutes of Li evaporation, along the $K$-point momentum cut indicated by the white line in the Fermi surface plot in (e). The Dirac cone Fermi surface was measured at this specific K point, and then replicated at the other K points by symmetry (note that high-symmetry points are here defined for the Brillouin zone of pristine graphene and not of $\sqrt{3}\times\sqrt{3}R30^{\circ}$ reconstructed Li-graphene, which is instead the notation in Ref. \onlinecite{Profeta2012}). The point at which the spectroscopic gap is studied is indicated by the shaded white circle. The Dirac point, already located below $E_F$ on pristine graphene due to the charge-transfer from the SiC substrate (a), further shifts to higher energies with Li evaporation (b). The presence of a single well-defined Dirac cone indicates a macroscopically uniform Li-induced doping. While no bands are present at the $\Gamma$-point on pristine graphene (c), spectral weight is detected on 10-minute Li-decorated graphene in (d) and (e). As illustrated in the 8\,K sheet carrier density plot versus Li deposition time in (f), which accounts for the filling of the $\pi^*$ Fermi surface, the spectral weight at $\Gamma$ is observed for charge densities $n_{2D}\!\gtrsim\!9\!\times\!10^{13}$cm$^{-2}$ (but completely disappears if the sample temperature is raised above $\sim\!50$\,K, and is not recovered on subsequent cooling; see also SI Appendix).}
\end{figure*}

ARPES measurements of the electronic dispersion of pristine and Li-decorated graphene at 8\,K, characterized by the distinctive Dirac cones at the corners of the hexagonal Brillouin zone [Fig.\,\ref{Fig:1}(e)], are shown in Fig.\,\ref{Fig:1}(a) and (b). Li adatoms electron-dope the graphene sheet via charge-transfer doping, leading to a shift of the Dirac point to higher binding energies. As evidenced by the evolution of the graphene sheet carrier density in Fig.\,\ref{Fig:1}(f), this trend begins to saturate after several minutes of Li deposition. Concomitantly, we observe the emergence of new spectral weight at the Brillouin zone centre [see Fig.\,\ref{Fig:1}(e) and the comparison of the $\Gamma$-point ARPES dispersion for pristine and 10 minute Li-decorated graphene in Fig.\,\ref{Fig:1}(c,d)]. 
The origin of this spectral weight is probably the Li-2$s$ band expected for this system \cite{Profeta2012}, superimposed with the folded graphene bands due to a Li superstructure, as observed in Li and Ca bulk GIC systems \cite{Sugawara2008,Yang2014}. This spectral weight, which disappears above $\sim\!50$\,K and is not recovered on subsequent cooling (see SI Appendix), is associated with the strong enhancement of electron-phonon coupling to be discussed later.

\begin{figure*}[t!]
\includegraphics[width=0.96\textwidth]{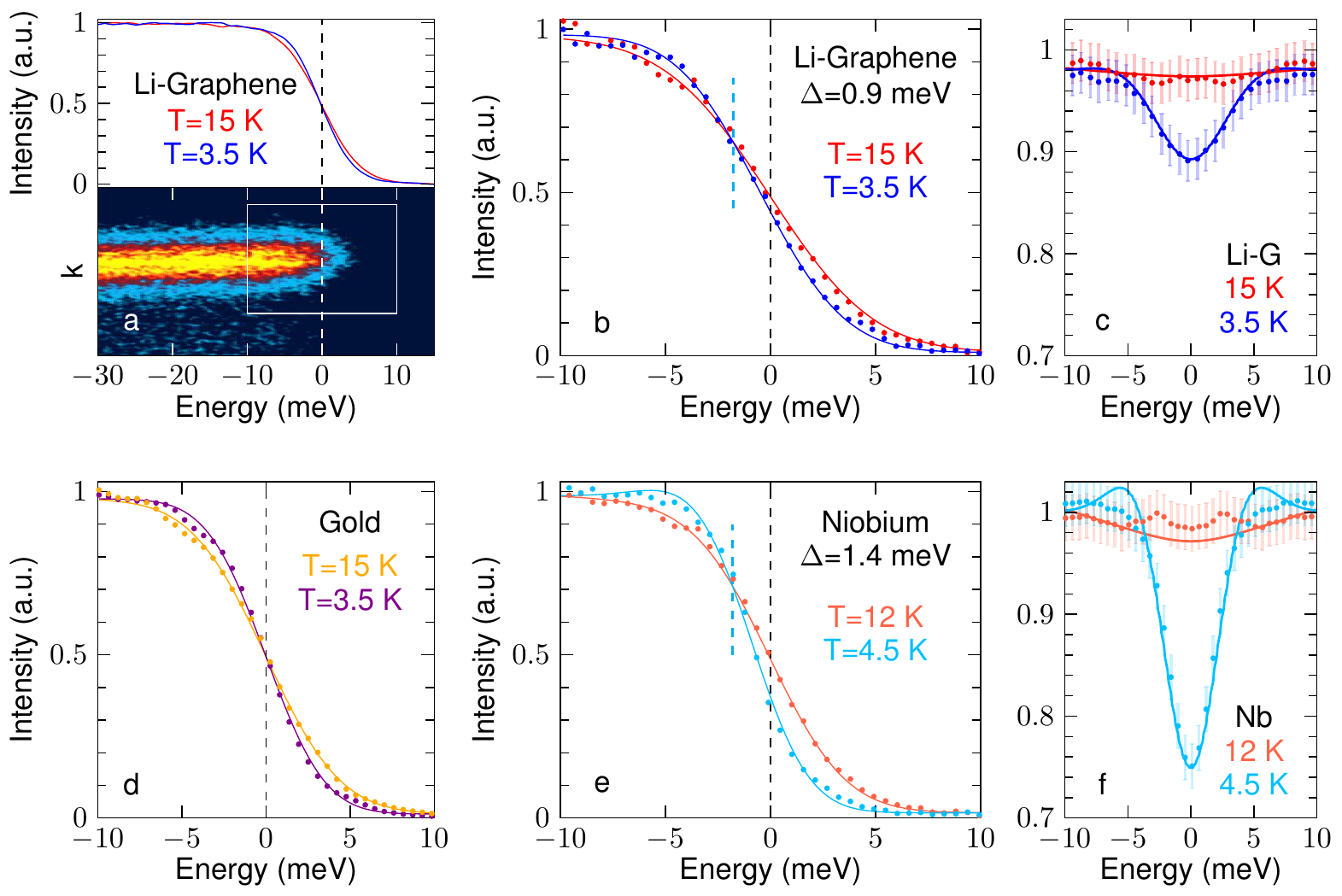}
\caption{\label{Fig:3} \textbf{Spectroscopic observation of a pairing gap in Li-decorated graphene.} (a) Dirac dispersion from 10-minute Li-decorated graphene measured at 15 and 3.5\,K, at the k-space location indicated by the white circle in Figs.\,\ref{Fig:1}(e) and \ref{Fig:2}(e); the temperature dependence is here evaluated for EDCs integrated in the 0.1\,\AA$^{-1}$ momentum region about $k_F$ shown by the white box (bottom), with the only changes occurring near $E_F$ (top). While Au spectra (d) cross at $E_f$ as described by the Fermi-Dirac distribution, the crossing point of the Li-graphene spectra (b) is shifted away from $E_F$ (cyan dashed line), due to the pull-back of the leading edge at 3.5\,K. A fit to the Dynes gap equation (see Methods) yields a gap $\Delta\!\simeq\!0.9$\,meV at 3.5\,K (and 0\,meV at 15\,K). The superconducting gap opening is best visualized in the symmetrized data in (c), i.e. by taking $I(\omega)\!+\!I(-\omega)$ which minimizes the effects of the Fermi function even in the case of finite energy and momentum resolutions \cite{Norman:1998,Mesot:2001} [blue and red symbols in (c) represent the smoothed data, while the light shading gives the root-mean-square deviation of the raw data]. The qualitatively similar behaviour observed on polycrystalline niobium -- and returning a superconducting gap $\Delta\!\simeq\!1.4$\,meV -- is shown in (e) and (f).}
\end{figure*}

\begin{figure*}[t!]
\includegraphics[width=\textwidth]{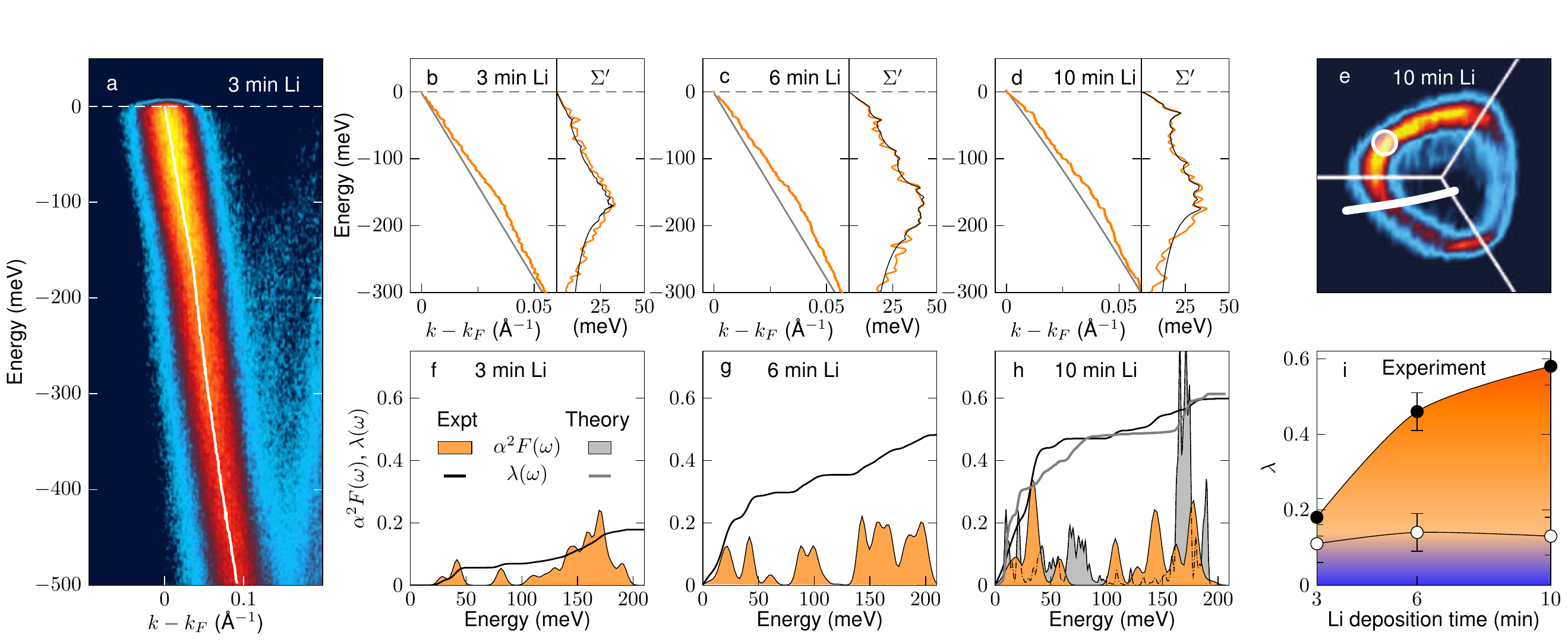}
\caption{\label{Fig:2} \textbf{Analysis of electron-phonon coupling in Li-decorated graphene.} (a) Dirac dispersion from 3-minute Li-decorated graphene, along the $k$-space cut indicated in the Fermi surface plot in (e), exhibiting kink anomalies due to electron-phonon coupling (white line: MDC dispersion). (b-d) MDC dispersion and bare-band obtained from the self-consistent Kramers-Kronig bare-band fitting (KKBF) routine \cite{Veenstra2010,Veenstra2011}, for several Li coverages (see Methods and SI Appendix); the real part of the self-energy $\Sigma^{\prime}$ is shown in the side panels (orange: $\Sigma^{\prime}$ from KKBF routine analysis; black: $\Sigma^{\prime}$ corresponding to the Eliashberg function presented below). (f-h) Eliashberg function $\alpha^2F(\omega)$ from the integral inversion of $\Sigma^{\prime}(\omega)$ \cite{Shi2004}, and electron-phonon coupling constant $\lambda\!=\!2\int\!d\omega\,\alpha^2F(\omega)/\omega$ (see Methods and SI Appendix); in (h) the theoretical result from Ref.\,\onlinecite{Profeta2012} for a LiC$_6$ monolayer are also shown (gray shading). (i)  Experimentally-determined contribution to the total electron-phonon coupling (black symbols) from: phonon modes in the energy range 100-250\,meV (blue shading, white symbols), and 0-100\,meV (orange shading); the coupling of low-energy modes strongly increases with Li coverage.}
\end{figure*}

Next we use high-resolution, low-temperature ARPES to search for the opening of a temperature-dependent pairing gap along the $\pi^*$-band Fermi surface, as a direct spectroscopic signature of the realization of a superconducting state in monolayer LiC$_6$. To increase our experimental sensitivity, as illustrated in Fig.\,\ref{Fig:3}(a) and following the approach introduced for FeAs \cite{Ev2009} and cuprate \cite{Reber2012} superconductors, we perform an analysis of ARPES energy distribution curves (EDC) integrated in $dk$ along a one-dimensional momentum-space cut perpendicular to the Fermi surface. This also provides the added benefit that the integrated EDCs can be modelled in terms of a simple Dynes gap function \cite{Dynes1978} multiplied by a linear background and the Fermi-Dirac distribution, all convolved with a Gaussian resolution function (see Methods and in particular Eq.\,\ref{eq:gap}). As shown in Fig.\,\ref{Fig:3}(a) and especially \ref{Fig:3}(b) for data from the k-space location indicated by the white circle in Figs.\,\ref{Fig:1}(e) and \ref{Fig:2}(e), a temperature dependence characteristic of the opening of a pairing gap can be observed near $E_F$. The leading edge midpoint of the Li-graphene spectra moves away from $E_F$ [Fig.\,\ref{Fig:3}(b)] in cooling from 15 to 3.5\,K, at variance with the case of Au spectra crossing precisely at $E_F$ according to the Fermi-Dirac distribution [Fig.\,\ref{Fig:3}(d)]. When fit to Eq.\,\ref{eq:gap}, this returns a gap value $\Delta\!=\!0.9\!\pm\!0.2$\,meV at 3.5\,K (with $\Gamma\!\simeq\!0.09$\,meV \cite{gamma}). Given its small value compared to the experimental resolution, the gap opening is best visualized in the symmetrized data in Fig.\,\ref{Fig:3}(c), which minimizes the effects of the Fermi function. Finally, we note that the gap appears to be anisotropic, and is either absent or below our detection limit along the $K\!-\!M$ direction (see Fig.\,4 in SI Appendix).

The detection of a temperature-dependent anisotropic gap at the Fermi level with a leading-edge profile described by the Dynes function -- with its asymmetry about $E_F$ and associated transfer of spectral weight to just below the gap edge -- is suggestive of a {\it superconducting pairing gap}. 
The phenomenology would in fact be very different in the case of a Coulomb gap, typically observed in disordered semiconductors \cite{Coulomb,Coulomb2,Coulomb1}, due to the combination of disorder with long-range Coulomb interactions. This would lead to a rigid shift of the spectra leading edge, isotropic in momentum, and result in a vanishing of the momentum-integrated density of states at E$_F$.

Similarly, the observed gap is unlikely to have a charge density wave origin, since the gap is tied to the Fermi energy as opposed to a particular high-symmetry wavevector (the latter might occur at the M points, when graphene is doped all the way to the Van Hove singularities resulting in a highly-nested hexagonal Fermi surface, or at the K points in the case of a $\sqrt{3}\,\times\sqrt{3}\,R\,30^{\circ}$ reconstruction leading to a Dirac-point gap). Finally, we note that these measurements do not allow us to speculate on the precise symmetry of the gap along a single Dirac-cone Fermi surface, nor on the relative phases of the gap on the six disconnected Fermi pockets. As such, our results do not rule out any of the recent proposals for a possible unconventional superconducting order parameter (see for example Refs.\,\onlinecite{gapsym1,Nandkishore2012,gapsym2}).

To further explore the nature of the gap observed on Li-decorated graphene (and also demonstrate our ability to resolve a gap of the order of 1\,meV), in Fig.\,\ref{Fig:3}(e,f) we show as a bench-mark comparison the analogous results from a bulk, polycrystalline niobium sample -- a known BCS superconductor with $T_c\!\simeq\!9.2K$. The Dynes fit of the integrated EDCs Fermi edge in Fig.\,\ref{Fig:3}(e) determines the gap to be $\Delta\!=\!1.4\!\pm\!0.2$ meV (with $\Gamma\!\simeq\!0.14$\,meV \cite{gamma}), in excellent agreement with reported values \cite{Carbotte1990}. Although the leading edge shift [Fig.\,\ref{Fig:3}(e)] and the dip in the symmetrized spectra [Fig.\,\ref{Fig:3}(f)] are more pronounced than for Li-graphene owing to the larger gap, the behaviour is qualitatively very similar. This provides additional support to the superconducting origin of the temperature-dependent gap observed in Li-decorated graphene.

If this is indeed a superconducting gap, the responsible mechanism may likely be electron-phonon coupling, as predicted by theory for monolayer Li-graphene \citep{Profeta2012} and also seen experimentally for bulk GIC CaC$_6$ \cite{Yang2014}. In direct support of this scenario, we present a detailed analysis of the graphene $\pi^{\ast}$ bands in Fig.\,\ref{Fig:2}, demonstrating that the Li-induced enhancement of the electron-phonon coupling is indeed sufficient to stabilize a low-temperature superconducting state. Graphene doped with alkali adatoms always shows a strong kink in the $\pi^*$ band dispersion at a binding energy of about 160\,meV \cite{Fedorov2014}. For the Li-graphene studied here, this is seen in the momentum-distribution curve (MDC) dispersions and corresponding real part of the self-energy $\Sigma^{\prime}$ in Fig.\,\ref{Fig:2}(b-d). This structure stems from the coupling to carbon in-plane (C$_{xy}$) phonons \cite{Profeta2012,Calandra2005a}. Despite the apparent strength of this kink, the interaction with these phonon modes contributes little to the overall coupling parameter due to their high energy (note that $\omega$ appears as a weighting factor in the integral calculation of $\lambda$ - see Methods). As illustrated by the white symbols in Fig.\,\ref{Fig:2}(i), the contribution to $\lambda$ from these high-energy (100-200\,meV) modes is determined to be $0.14\!\pm\!0.05$, and it remains approximately constant for all Li coverages studied here. This value is, however, too small to stabilize a  superconducting state in this system \citep{Profeta2012,Fedorov2014}.

With increasing Li coverage and the appearance of the spectral weight at $\Gamma$, significant modifications to the low-energy part of the dispersion ($\lesssim \!100$\,meV) become apparent [Fig.\,\ref{Fig:2}(b-d)]. With 10 minutes of Li deposition [Fig.\,\ref{Fig:2}(d)], an additional kink is visible at a binding energy of approximately 30\,meV, along with the associated peak in the real part of the self-energy $\Sigma^{\prime}$. The extracted (see Methods) Eliashberg functions and energy-resolved $\lambda(\omega)$ in Fig.\,\ref{Fig:2}(f-h) show that, at high Li coverage, phonon modes at energies below 60\,meV are coupling strongly to the graphene electronic excitations.  The phonon modes in this energy range are of Li in-plane (Li$_{xy}$) and C out-of-plane (C$_z$) character \cite{Profeta2012,Calandra2005a}. This is in agreement with predictions \cite{Profeta2012}, as shown by the direct comparison between theory and experiment in Fig.\,\ref{Fig:2}(h) \cite{eliashberg}. As for the total electron-phonon coupling $\lambda$ for each coverage [black symbols in Fig.\,\ref{Fig:2}(i)], our values measured on the $\pi^*$-band Fermi surface at an intermediate location between $\Gamma\!-\!K$ and $K\!-\!M$ directions [Fig.\,\ref{Fig:2}(e)] provide an effective estimate for the momentum-averaged coupling strength \citep{strength}. Remarkably, the value $\lambda\!=\!0.58\pm0.05$ observed at the highest Li coverage [Fig.\,\ref{Fig:2}(i)] is comparable with $\lambda\!=\!0.61$ predicted for monolayer LiC$_6$ \cite{Profeta2012} as well as $\lambda\!\simeq\!0.58$ observed for bulk CaC$_6$ \citep{Valla2009} --  it is thus large enough for inducing superconductivity in Li-decorated graphene. It is also significantly larger than the momentum-averaged results previously reported for both Li and Ca deposition on monolayer graphene ($\lambda\!\simeq\!0.22$ and 0.28, respectively \citep{Fedorov2014}). We note that achieving such a large  $\lambda$ value is critically dependent on the presence of the spectral weight observed at $\Gamma$ when Li is deposited on graphene at low temperatures, presumably forming an ordered structure on the surface and not intercalating. As shown in the SI Appendix, we find $\lambda\!=\!0.13\pm0.05$ after the same sample is annealed at 60\,K for several minutes, destroying the Li order and associated $\Gamma$ spectral weight.

Taken together, our ARPES study of Li-decorated monolayer graphene provides the first evidence for the presence of a temperature-dependent pairing gap on part of the graphene-derived $\pi^*$ Fermi surface. The detailed evolution of the density of states at the gap edge, as well as the phenomenology analogous to the one of known superconductors such as Nb -- as well as CaC$_6$ and NbSe$_2$, which also show a similarly anisotropic gap around the $K$ point \cite{Sanna2007,Sanna2012,Gonnelli2008,NbSe2_1,NbSe2_2} -- indicate that the pairing gap observed at 3.5\,K in graphene is most likely associated with superconductivity. Based on the BCS gap equation, $\Delta\!=\!3.5\,k_b\,T_c$, this suggests that Li-decorated graphene is superconducting with $T_c\!\simeq\!5.9$\,K, remarkably close to the value of 8.1\,K found in density-functional theory calculations \citep{Profeta2012}. This constitutes the first experimental realization of superconductivity in graphene -- the most prominent electronic phenomenon still missing among the remarkable properties of this single layer of carbon atoms.

\section{Acknowledgements}

We gratefully acknowledge D.A. Bonn, S.A. Burke, M. Calandra, A. Chubukov, E.H. da Silva Neto, J.A. Folk, M. Franz, P. Hofmann, A.F. Morpurgo, G. Profeta, G.A. Sawatzky, and S. Ulstrup for valuable discussions, and P. Trochtchanovitch and M. O'Keane for technical assistance. This work was supported by the Max Planck - UBC Centre for Quantum Materials, the Killam, Alfred P. Sloan, Alexander von Humboldt, and NSERC's Steacie Memorial Fellowship Programs (A.D.), the Canada Research Chairs Program (A.D.), the NSERC PDF Scholarship (S.Z.), NSERC, CFI, and CIFAR Quantum Materials.
\section{Methods}

\noindent\textbf{Sample preparation.}\,Epitaxial graphene monolayers with a carbon buffer layer were grown under argon atmosphere on hydrogen-etched 6H-SiC(0001) substrates, as described in Ref.\,\onlinecite{Forti2014}. The samples were annealed at 500$^\circ$C and 8$\times$10$^{-10}$\,Torr for 1 hour, immediately prior to the ARPES measurements. Lithium adatoms were deposited from a commercial SAES alkali metal source, with the graphene samples held at a temperature of 8\,K. Bulk Nb polycrystalline samples, with $T_c\!=\!9.2$\,K, were cleaved in the ARPES chamber prior to the experiments.

\noindent\textbf{ARPES experiments.}\,The measurements were performed at UBC with $s$-polarized 21.2\,eV photons on an ARPES spectrometer equipped with a SPECS Phoibos 150 hemispherical analyzer, a SPECS UVS300 monochromatized gas discharge lamp, and a 6-axes cryogenic manipulator that allows controlling the sample temperature between 300 and 3.5\,K, with accuracy $\pm\!0.1$\,K. Band and Fermi surface mapping, as well as the study of electron-phonon coupling, were performed at 8\,K with energy and angular resolution set to 15 meV and 0.01\,\AA$^{-1}$, respectively. For the measurements of the superconducting gaps, energy and angular resolution were set to 6 meV and 0.01\,\AA$^{-1}$, while the sample temperature was varied between 3.5 and 15\,K. During the ARPES measurement the chamber pressure was better than 4$\times$10$^{-11}$\,Torr.

\noindent\textbf{Electron-phonon coupling analysis.}\,The spectral function $A({\bf k},\omega)$ measured by ARPES \cite{Damascelli:reviewPS} provides information on both the single-particle electronic dispersion $\varepsilon^b_{\bf k}$ (the so-called `bare-band') as well as the quasiparticle self-energy $\Sigma({\bf k},\omega)\!=\!\Sigma^{\prime}({\bf k},\omega)\!+\!i\Sigma^{\prime\prime}({\bf k},\omega)$, whose real and imaginary parts account for the renormalization of electron energy and lifetime due to many-body interactions, including electron-phonon coupling. By fitting with a Lorentzian and a constant background the ARPES intensity profiles at constant energy $\omega=\tilde\omega$, known as momentum distribution curves (MDCs), one obtains the MDC dispersion defined by the peak maximum $k_m$ [plotted in Fig.\,\ref{Fig:2}(a-d)], as well as the corresponding half-width half-maximum (HWHM) $\Delta k_m$. The real and imaginary parts of the self-energy can then be defined as:
\begin{eqnarray}
\Sigma^{\prime}_{\tilde\omega}&=&\tilde\omega - \varepsilon^b_{k_m},
\nonumber \\ \Sigma^{\prime\prime}_{\tilde\omega}&=&-\Delta k_m
v^b_{k_m},
\label{eqn:mdc_params}
\end{eqnarray}
\noindent
(where $v^b_{k_m}$ is the bare-band velocity). To extract the self-energy and dispersion without any a priori knowledge of the bare-band, we use the self-consistent Kramers-Kronig bare-band fitting (KKBF) routine from Ref.\,\onlinecite{Veenstra2010} and\,\onlinecite{Veenstra2011} (see also SI Appendix).

As for the dimensionless $k$-resolved electron-phonon coupling constant discussed in the paper and in particular in Fig.\,\ref{Fig:2}(f-i), this is formally defined as \cite{Grimvall}: 
\begin{equation}
\lambda_{{\bf k}}(\omega)=\,2\int_0^{\omega}\!d\omega^{\prime}\, \frac{\alpha^2F({\bf k},\omega^{\prime})}{\omega^{\prime}}\, ,
\label{eqn:lambda}
\end{equation}
where $\alpha^2F({\bf k},\omega)$ is the Eliashberg function, i.e. the phonon density of states weighted by the electron-phonon coupling strength \cite{Grimvall}. The latter is related to the real part of the self-energy $\Sigma^{\prime}({\bf k},\omega)$ via the integral relation:
\begin{equation}
\Sigma^{\prime}({\bf k},\omega)\!=\!\int_0^{\infty}\!d\omega^{\prime}\, \alpha^2F({\bf k},\omega^{\prime})\,K\left(\frac{\omega}{kT},\frac{\omega^{\prime}}{kT}\right)\, ,
\label{eqn:sigma}
\end{equation}
where $K(y,y^{\prime})\!=\!\int_{-\infty}^{+\infty}\!dx\,f(x-y)\,2y^{\prime}/(x^2-y^{\prime \, 2})$ and $f(x-y)$ is the Fermi-Dirac distribution. The momentum resolved $\alpha^2F({\bf k},\omega)$ function plotted in Fig.\,\ref{Fig:2}(f-h) can then be extracted from the real part of the self-energy $\Sigma^{\prime}({\bf k},\omega)$ probed by ARPES via the integral inversion procedure described in Ref.\,\onlinecite{Shi2004}. Ultimately, by means of Eq.\,\ref{eqn:lambda}, this also allows the calculation of the electron-phonon coupling constant shown in Fig.\,\ref{Fig:2} (see also SI Appendix).

\noindent\textbf{Superconducting gap fitting.}\,As discussed in detail for the case of FeAs and cuprate superconductors in Ref.\,\onlinecite{Ev2009} and \,\onlinecite{Reber2012}, respectively, when evaluating the opening of a superconducting gap based on the EDCs integrated in $dk$ along a one-dimensional momentum-space cut perpendicular to the Fermi surface (such as those presented in Fig.\,\ref{Fig:3}), one can make use of the following formula:
\begin{equation}
I_{\!\int\!\!dk}(\omega)\!=\!\left[f(\omega,T)\,(a\!+\!b\,\omega) \left|{\rm Re}\frac{\omega-i\Gamma}{\sqrt{(\omega-i\Gamma)^2-\Delta^2}}\right|\,\right]\!\otimes R_{\omega}\, .
\label{eq:gap}
\end{equation}
This corresponds to the Dynes gap function (i.e., the Bardeen-Cooper-Schrieffer density of states with a superconducting gap $\Delta$, broadened by the pair-breaking scattering rate $\Gamma$ \cite{Dynes1978}), multiplied by a linear background (with parameters $a$ and $b$) and by the Fermi-Dirac distribution $f(\omega,T)$, all convolved with a Gaussian function $R_{\omega}$ accounting for the experimental energy resolution (owing to the integration of the ARPES intensity in $dk$, this analysis is unaffected by momentum resolution \cite{Ev2009}). The high and low temperature ARPES data in Fig.\,\ref{Fig:3} are fitted simultaneously using the above equation, according to the following additional  considerations: since the integrated EDCs were observed not to change outside of the gap region, the linear background is constrained to be the same for the above and below $T_c$ measurements; the temperatures are fixed to the known measured values (with accuracy $\pm\!0.1$\,K); the Fermi energy  $E_F$ and energy resolution are determined from fitting the high temperature data from either Nb and Li-graphene during each measurement, and independently verified from measurements on polycrystalline gold.


\begin{thebibliography}{10}
\bibitem{gamma}
\bibinfo{title}{{{The parameter $\Gamma$ in the Dynes fitting function is not
  treated as a free fitting parameter as the broadening of the coherence peaks
  and filling in of the gap is dominated by the experimental energy resolution.
  However, setting this parameter to small, realistic values ($\Gamma\sim
  0.1\Delta$) improves the fit at the centre of the gap (i.e. at E=0 in the
  symmetrized data), without affecting the value of the gap itself\!\!}}} .
\bibitem{eliashberg}
\bibinfo{title}{{{As for the theoretical and experimental Eliashberg functions
  $\alpha^2F(k,\omega)$ in Fig.\,\ref{Fig:2}(h), the agreement may at first
  glance appear not as good as the one for $\lambda(\omega)$. We note however
  that, in this regard, the relevant information is in the macroscopic energy
  distribution of the $\alpha^2F(k,\omega)$ weight, rather than in its detailed
  structure\!\!}}} .
\end{thebibliography}
\end{document}